\documentclass[10pt,twocolumn,superscriptaddress,showpacs,aps,amsmath,amssymb,nofootinbib,prl]{revtex4-1}
\pdfoutput=1

\usepackage[utf8]{inputenc}

\usepackage{bm}
\usepackage{textcomp}
\usepackage{bbm}
\usepackage{enumerate}
\usepackage{multirow}

\usepackage{tikz}
\usetikzlibrary{calc}

\usepackage{hyperref}
\hypersetup{%
pdftitle={Finding the ground state of the Hubbard model by variational methods on a quantum computer with gate errors},%
pdfauthor={Jan-Michael Reiner, Frank-Wilhelm Mauch, Gerd Sch\"on, and Michael Marthaler},%
colorlinks=true,%
linkcolor=blue,%
urlcolor=blue,%
citecolor=blue
}

\newcommand{\e}{\mathrm{e}}

\newcommand{\gate}[1]{\texttt{#1}}
\renewcommand{\sp}[1]{\sigma^+_{#1}} 
\newcommand{\sm}[1]{\sigma^-_{#1}}

\renewcommand{\vec}[1]{{\bm #1}}

\begin{document}

\title{Finding the ground state of the Hubbard model by variational methods on a quantum computer with gate errors}

\author{Jan-Michael Reiner}
\affiliation {Institut f\"ur Theoretische Festk\"orperphysik, Karlsruhe Institute of Technology (KIT), 76131 Karlsruhe, Germany}
\affiliation {HQS Quantum Simulations, c/o CyberForum Service GmbH, Haid-und-Neu-Straße 18, 76131 Karlsruhe, Germany}

\author{Frank Wilhelm-Mauch}
\affiliation {Theoretical Physics, Saarland University, 66123 Saarbr\"ucken, Germany}

\author{Gerd Sch\"on}
\affiliation {Institut f\"ur Theoretische Festk\"orperphysik, Karlsruhe Institute of Technology (KIT), 76131 Karlsruhe, Germany}
\affiliation {Institute of Nanotechnology, Karlsruhe Institute of Technology (KIT), 76021 Karlsruhe, Germany}

\author{Michael Marthaler}
\affiliation {Institut f\"ur Theoretische Festk\"orperphysik, Karlsruhe Institute of Technology (KIT), 76131 Karlsruhe, Germany}
\affiliation {HQS Quantum Simulations, c/o CyberForum Service GmbH, Haid-und-Neu-Straße 18, 76131 Karlsruhe, Germany}
\affiliation {Theoretical Physics, Saarland University, 66123 Saarbr\"ucken, Germany}

\date{\today}

\begin{abstract}
A key goal of digital quantum computing is the simulation of fermionic systems such as molecules or the Hubbard model. Unfortunately, for present and near-future quantum computers the use of quantum error correction schemes is still out of reach. Hence, the finite error rate limits the use of quantum computers to algorithms with a low number of gates. The variational Hamiltonian ansatz (VHA) has been shown to produce the ground state in good approximation in a manageable number of steps. Here we study explicitly the effect of gate errors on its performance. The VHA is inspired by the adiabatic quantum evolution under the influence of a time-dependent Hamiltonian, where the -- ideally short -- fixed Trotter time steps are replaced by variational parameters. The method profits substantially from quantum variational error suppression, e.g., unitary quasi-static errors are mitigated within the algorithm. We test the performance of the VHA when applied to the Hubbard model in the presence of unitary control errors on quantum computers with realistic gate fidelities.
\end{abstract}


\maketitle

\section{Introduction}

Simulating systems of strongly correlated electrons, such as the iconic Hubbard model, is a key goal of condensed matter physics. But important effects, such as as high-$T_C$ superconductivity or detailed magnetic properties still pose serious computational challenges. The hope is that digital quantum computers or quantum simulators would bring the needed progress. The Hubbard model and spin models have been studied in several proposals and experiments, e.g., with ultra-cold gases~\cite{Cold_Gases_Simulator,Cirac_Cold_gases,Fermi_Sea_Heidelberg} and trapped ions~\cite{350_Spin_simulator,Cirac_PRL,Quantum_Magnet_Schaetz,Ion_Simulator}. These experiments can be considered \emph{analog} simulations, where the system to be studied is recreated by a well controllable artificial one. The goal is to simulate systems which are beyond the reach of classical computations. But so far classical simulations can match all existing fermionic analog simulators, and the experiments on the fermionic Hubbard model -- while representing impressive technological advances -- are still at the proof-of-principle state. One of the problems is that analog simulators based on fermions are limited to high temperatures as compared to the intrinsic coupling strengths~\cite{Temperature_Fermionic}.

In recent years, systems with increasing numbers of high-fidelity and fully controllable Josephson qubits have become available, and they were integrated in a single processor. This opens the perspective of simulating the Hubbard model, e.g., its time evolution or correlation functions, using a \emph{gate-based} approach~\cite{Hubbard_Frank,Frank_Hubbard_2,Efficent_Hubbard}. Qubits with fidelities at the threshold for the implementation of quantum error correction have been demonstrated~\cite{Marinis_Threshold,Blatt_error_correction}. However, for the near-term prospects the number of qubits required for full quantum error correction is prohibitively large~\cite{Devitt_error_correction,Fowler_Surface_Code,Scalable_Hensinger}. Hence, for meaningful near-term applications it is crucial to estimate the effects of errors~\cite{Jan_Noise_Paper}. For certain situations, methods to verify the performance of quantum simulators with errors have been suggested~\cite{Certification_Eisert,Certification_Marthaler}, and some proposals for error reduction exist~\cite{Correction_Marthaler,Correction_IBM}.

For one of the important goals, the simulation of the ground state of a quantum system, it has been suggested and demonstrated in few-qubit experiments that variational algorithms require only a relatively low number of gates and, in addition, variational methods intrinsically suppress the impact of errors ~\cite{Theoretische_Noise_Reduction,Experimental_Noise_Reduction,VQE_Paper}. In general, variational approaches apply a unitary operator to an initial state $| \psi_0 \rangle$ that is easy to prepare. The unitary operator $U(\vec \theta)$ depends on a set of parameters $\vec{\theta}$ that is varied to minimize the energy
\begin{equation}\label{eq:vha-energy}
E(\vec{\theta})= \langle \psi_0 | U^{\dag}(\vec{\theta}) H U(\vec{\theta}) | \psi_0 \rangle
\end{equation}
where $H$ is the Hamiltonian of the system of interest.

In this paper we study explicitly the effect of gate errors on a variational algorithm for finding the ground state of the Hubbard model. We will use a specific variational ansatz, namely the variational Hamiltonian ansatz (VHA)~\cite{VHA_Troyer_Paper}. It is inspired by the adiabatic ground state evolution as explained in more detail below. Specifically we address the following questions: How close can $E(\vec \theta)$ get to the ground state energy $E_\mathrm{g}$ of the Hamiltonian $H$, and how close can $U(\vec \theta) |\psi_0\rangle$ approximate the true ground state of $H$, if gate errors occur during the implementation of the unitary operator $U(\vec{\theta})$.

Generally the specific nature of gate errors is not known, therefore we work with a simple but representative model. Every gate can be interpreted as a rotation of the qubit register. In our model gate errors are modelled as over-rotations (or under-rotations). As discussed in earlier work~\cite{Jan_Noise_Paper} the over-rotation angle $\delta \varphi$ can be related to the minimal gate fidelity $\mathcal{F}_\mathrm{min}$ of the gate via $\mathcal{F}_\mathrm{min} = \cos(\delta \varphi)$.\footnote{Note that often the square of $\mathcal{F}_\mathrm{min}$ is called (minimal) gate fidelity. Our definition, relating $\mathcal{F}_\mathrm{min}$ to the magnitude of $\delta \varphi$, relates it equally to what is known as the Bures angle, associated with the Bures metric (i.e.\ the over-rotation angle magnitude and the Bures angle coincide).} Because of the vanishing slope of the cosine at the maximum (i.e., $|\delta\varphi| = \arccos(\mathcal{F}_\mathrm{min}) \approx \sqrt{2(1-\mathcal{F}_\mathrm{min})}$ for $\mathcal{F}_\mathrm{min} \approx 1$) gate fidelities need to get very close to $100\,\%$ to significantly limit the magnitude of the over-rotations.

Below we also compare the VHA to the adiabatic state preparation based on the Trotter expansion. We find that the VHA produces a better approximation to the ground state with far fewer steps, and therefore gates, than adiabatic state preparation. For adiabatic state preparation, even for weak gate errors, upon increasing the number of steps, the states created have decreasing overlap with the actual target ground state. In contrast, the VHA achieves high overlap with the exact ground state; even with gate errors. This is due to the error mitigation capabilities of variational approaches. For the (still small-size) Hubbard model considered as an example in the following we find that a gate fidelity of $\mathcal{F}_\mathrm{min} = 99.9\,\%$ is sufficient for a meaningful simulation.

\section{From adiabatic evolution to the variational Hamiltonian ansatz (VHA)}

The unitary operator $U(\vec \theta)$ of the variational Hamiltonian ansatz is based on the Hamiltonian itself: The different terms of $H$ are separated and grouped into $N$ sub-operators $H_1, \ldots, H_N$ such that $H = \sum_{\alpha=1}^N H_\alpha$. The unitary operator for the VHA with $n$ steps is
\begin{align}\label{eq:vha-evolution}
U(\vec \theta) = \prod_{k=1}^n \prod_{\alpha=1}^N \e^{\mathrm i \theta_{\alpha,k} H_\alpha},
\end{align}
where $\vec \theta$ collects all the variational parameters $\theta_{\alpha,k}$. The optimization criterion is the minimization of the energy expectation value~\eqref{eq:vha-energy} of the final state $|\psi_\mathrm{f}\rangle = U(\vec \theta) |\psi_0\rangle$ with respect to the $n \cdot N$ variation parameters $\vec \theta$ (the ground state is, per definition, the state with minimal energy).

The ansatz~\eqref{eq:vha-evolution} is inspired by the adiabatic time evolution under the influence of the Hamiltonian $H = H_0 + V$ composed of, e.g., a non-interacting part $H_0$ and the interaction $V$. If the interaction is turned on slowly on the time scales given by the inverse energy scales of the Hamiltonian, the initial ground state $| \psi_0 \rangle$ of $H_0$ develops adiabatically into the ground state $| \psi_\mathrm{g} \rangle$ of $H$. To simulate this evolution in a Trotter expansion the time $\tau$ of the evolution is divided into a large number $n$ of Trotter time steps $\tau/n$, each shorter than the inherent time scales, leading to
\begin{align}\label{eq:adiabatic-evolution}
U_\mathrm{ad} = \prod_{k=1}^n \e^{-\mathrm i \frac{\tau}{n} H_0} \e^{-\mathrm i \frac{\tau}{n} \frac{k}{n} V} \, .
\end{align}
During each of the short time steps one further decomposes the Hamiltonian into sub-operators. In a simulation using an available quantum computer the sub-operators are chosen such that the short time evolution can be realized by the available gate operations.

The similarity between the operators~\eqref{eq:vha-evolution} and~\eqref{eq:adiabatic-evolution} justifies the expectation that the VHA can produce the evolution from a ground state $|\psi_0\rangle$ of the non-interacting system to the ground state $|\psi_\mathrm{g}\rangle$ of the full Hamiltonian. In addition, by introducing variational parameters the VHA can deviate from the adiabatic path and follow, through optimization, a more efficient one. Having a more efficient evolution via VHA allows for greatly reducing the necessary number of steps $n$, as compared to the number of Trotter steps in an adiabatic evolution, while still achieving high accuracy.\footnote{Note that to a certain degree the ansatz also helps us to cope with the so called Trotter error, which is the error introduced by decomposing the time evolution operator using a finite number of Trotter steps and grows as the number of steps is reduced.} Moreover, by optimizing the variational parameters one also mitigates the error introduced by faulty gates, an effect which had been termed \emph{variational error suppression}.

\section{Model Hamiltonian, its decomposition, and mapping to qubits}

The model system we investigate in this paper is the Hubbard Hamiltonian of spin-$\frac{1}{2}$ fermions
\begin{align}\label{eq:Hamiltonian}
H = -t \sum_{\substack{\langle j, j' \rangle \\ s=\uparrow,\downarrow}} (c^\dagger_{j,s} c^{\phantom \dagger}_{j',s} + c^\dagger_{j',s} c^{\phantom \dagger}_{j,s}) + U \sum_j c^\dagger_{j,\uparrow} c^{\phantom \dagger}_{j,\uparrow} c^\dagger_{j,\downarrow} c^{\phantom \dagger}_{j,\downarrow},
\end{align}
with hopping amplitude $t$ between nearest neighbors $\langle j, j' \rangle$, on-site energy $U$, and $c^{(\dagger)}_{j,s}$ being the annihilation (creation) operator of a fermion on site $j$ with spin $s$. In the following we consider two-dimensional square lattices and focus on the parameter values $U = 2t$ with repulsive on-site interaction, $U>0$.

\begin{figure}
\centering
\includegraphics[width=.75\columnwidth]{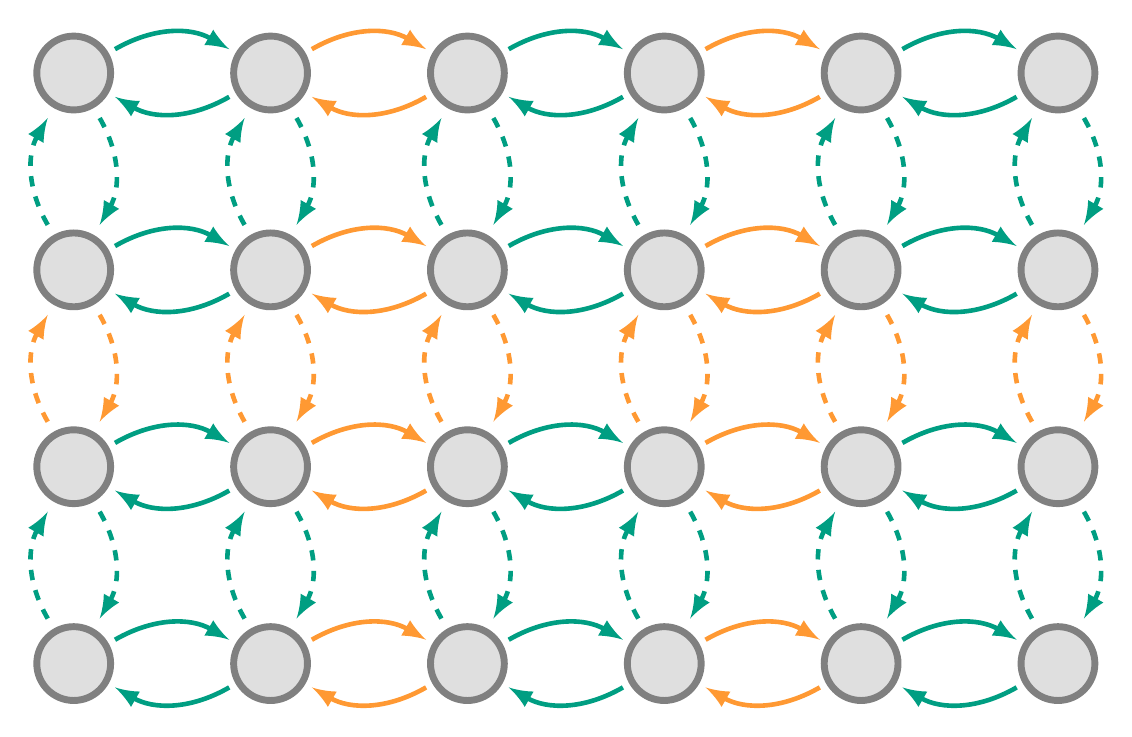}
\caption{%
Sketch of the 2D Hubbard model on a square lattice. The arrows indicate the hopping terms between neighboring sites. For the simulation we divide them into four sets: First in horizontal (solid lines) and vertical (dashed lines) directions. Then we subdivide for each direction into even (green) and odd (orange) terms. Summing up all hopping terms for each of the four sets yields the sub-Hamiltonians $H_1, \ldots, H_4$. Note that the hopping terms within a set commute among each other.
}
\label{fig:hopping-sketch}
\end{figure}

For the implementation of the variational unitary operation of Eq.~\eqref{eq:vha-evolution} we separate the Hamiltonian into $N=5$ parts: The non-interacting part is split into four terms, $H_1, \ldots, H_4$, as illustrated by Fig.~\ref{fig:hopping-sketch}. We distinguish between hopping terms in horizontal and vertical direction and for each direction we group even and odd terms, i.e., every other term in each direction of the 2D system. The on-site interaction terms are collected in $H_5$. Note that all terms collected within one $H_\alpha$ commute among each other. Hence, the execution of an exponential $\e^{\mathrm i \theta_{\alpha,k} H_\alpha}$ in Eq.~\eqref{eq:vha-evolution} can be performed exactly by sequentially applying the gates that account for the individual terms, without introducing an error associated with the Trotter expansion.

As written explicitly above we introduce a manageable number of 5 variational parameters per step and -- as the example below shows -- we reach high-quality results already for 10 steps or less.

The mapping onto a qubit system is performed via the Jordan-Wigner transformation. Gates performing on-site interaction terms are realized by $ZZ$-type interactions between qubits, hopping terms require $XX+YY$ interactions. In addition the transformation introduces \emph{Jordan-Wigner strings} which are implemented by additional chains of controlled $Z$ gates. A precise description of the mapping and gate sequence of the algorithm is presented in the \hyperref[appx:gate-sequence]{Appendix}.

\section{Error model and procedure}

The goal is to prepare the ground state $|\psi_\mathrm{g} \rangle$ of the Hubbard model~\eqref{eq:Hamiltonian} on a quantum computer. We start from the ground state $|\psi_0\rangle$ of the non-interacting model ($U=0$),\footnote{Note that the non-interacting system has a degenerate ground state. We hand-picked the correct ground state $|\psi_0\rangle$ that evolves towards $|\psi_\mathrm{g}\rangle$ when performing the VHA or the adiabatic evolution. To do this, we analyzed the spectrum where we lifted the degeneracy by applying a small perturbation through on-site interactions of negligible strength.} which -- in principle -- can be prepared efficiently on a quantum computer~\cite{Hubbard_Model_State_Preparation}, and we apply the variational Hamiltonian ansatz in order to evolve this state towards $| \psi_\mathrm{f} \rangle$, which should be close to the ground state $|\psi_\mathrm{g}\rangle$. We modeled the algorithm and the quantum gates including the gate errors on a classical computer. For the system sizes considered, $|\psi_0\rangle$ and $|\psi_\mathrm{g}\rangle$ (without errors) can also be found exactly through classical numerical diagonalization. This allows us to test the quality of the results.

The unitary evolution is implemented by a gate based algorithm described in more detail in the \hyperref[appx:gate-sequence]{Appendix}. Each gate can be written in the form $\e^{\mathrm i \varphi A}$ with a real angle $\varphi$ and an operator $A$ composed of Pauli operators. Hence, it can be interpreted as a rotation of the quantum state. We model unitary gate errors by over-rotations $\delta \varphi$ (which may be positive of negative), such that the faulty gate reads $\e^{\mathrm i (\varphi + \delta \varphi) A}$. The magnitude of the random $\delta \varphi$ is given by the minimal gate fidelity $\mathcal{F}_\mathrm{min}$, where $\mathcal{F}_\mathrm{min} = \cos (\delta \varphi)$~\cite{Jan_Noise_Paper}. Performing a sequence of gates with random normally distributed over-rotations with zero mean and variance $\mathrm{Var}(\delta \varphi)$ one finds -- for weak over-rotations, i.e., fidelities close to one -- an averaged minimal gate fidelity $\overline{\mathcal{F}}_\mathrm{min} = 1 - \mathrm{Var}(\delta \varphi)^2/2$. When we introduce gate errors in the following, we assume a certain gate fidelity $\overline{\mathcal{F}}_\mathrm{min}$ and add random over-rotations to each gate according to the above relation. However, once the over-rotation for a specific gate is chosen, this value is kept constant during the consecutive stages of the optimization process. This accounts for quasi-static errors, which are considered an appropriate noise model for superconducting qubits, where the noise spectrum is dominated by low frequencies~\cite{Noise_measurments_Bylander}.

Finally, for given gate fidelities, we measure the performance of the VHA by evaluating the \emph{final state fidelity}. This quantity is defined as the absolute value of the overlap $|\langle \psi_\mathrm{g} | \psi_\mathrm{f} \rangle |$ between the exactly known ground state $|\psi_\mathrm{g} \rangle$ and the final state $|\psi_\mathrm{f} \rangle = U(\vec \theta) | \psi_0 \rangle$ of the VHA according to Eq.~\eqref{eq:vha-evolution}, after the optimization of the parameters $\vec \theta$.\footnote{Note that frequently the square of the overlap is denoted as state fidelity. We chose it such that it is consistent with our definition of the gate fidelity.}

\section{VHA versus Adiabatic evolution}

We first study the quality of the variational Hamiltonian ansatz, Eq.~\eqref{eq:vha-evolution}, in comparison to the adiabatic evolution, Eq.~\eqref{eq:adiabatic-evolution}. In both cases we start from the same initial state $|\psi_0\rangle$. For the comparison, the same number of steps $n$ (steps of the VHA or Trotter steps) and the same gate sequence is used (with appropriately different parameters). 

For low $n$, the gate sequence introduces an error while implementing the exponential $\e^{-\mathrm{i} \frac{\tau}{n} H_0}$ for the adiabatic evolution, since the summands of $H_0$ do not commute. But for the comparison with the VHA at equal gate count we did not introduce finer Trotter time steps. On the other hand, the time $\tau$ in the adiabatic evolution was optimized for the given number of Trotter steps $n$.

Table~\ref{tab:adiabatic-vs-vha} shows the results for a two-dimensional Hubbard lattice with $2\!\times\!2$ sites. We present the final state fidelity $|\langle \psi_\mathrm{g} | \psi_\mathrm{f} \rangle |$ in percent after performing, on one hand, the adiabatic evolution and, on the other hand, the VHA for different numbers of (Trotter) steps $n$ and different averaged minimal gate fidelities $\overline{\mathcal{F}}_\mathrm{min}$. The algorithm we used requires $20$ two-qubit gates per (Trotter) step, i.e., in total we perform from $40$ to $100$ two-qubit gates.

\begin{table}
\begin{tabular}{r r r r r r}
\hline\hline
\multicolumn{5}{c}{Adiab.~evo., $2\!\times\!2$}\\
& & \multicolumn{3}{c}{$\bm{\overline{\mathcal{F}}_\mathrm{min}\,[\%]}$}\\
						 & & \textbf{100.0} & \textbf{99.99} & \textbf{99.90}\\
& \textbf{2}				  & 98.87			& 98.73			 & 97.43\\
\multirow{2}{*}{$\bm n$} & \textbf{3}				  & 99.15			& 98.85			 & 96.28\\
& \textbf{4}				  & 99.23			& 98.95			 & 96.51\\
& \textbf{5}				  & 99.55			& 99.14			 & 95.62\\
\hline\hline
\end{tabular}
\begin{tabular}{r r r r r r}
\hline\hline
\multicolumn{6}{c}{VHA, $2\!\times\!2$}\\
		&				   & \multicolumn{4}{c}{$\bm{\overline{\mathcal{F}}_\mathrm{min}\,[\%]}$} \\
		&				   & \textbf{100.0} & \textbf{99.99} & \textbf{99.90} & $\bm{99.90^*}$\\
 		& \textbf{2}	   & 99.68			& 99.63			 & 99.24		  & 94.44\\
\multirow{2}{*}{$\bm n$} & \textbf{3}	    & 100.0			 & 99.95		  & 99.56		  & 93.54\\
		& \textbf{4}	   & 100.0			& 99.96			 & 99.68		  & 88.77\\
		& \textbf{5}	   & 100.0			& 99.98			 & 99.82		  & 83.69\\
\hline\hline
\end{tabular}
\caption{%
The final state fidelity $|\langle \psi_\mathrm{g} | \psi_\mathrm{f} \rangle |$ for a given initial state $|\psi_0\rangle$ in percent for different numbers of (Trotter) steps $n$ and different values of the average minimal gate fidelity $\overline{\mathcal{F}}_\mathrm{min}$ for a $2\!\times\!2$ Hubbard system. The left side shows the results of the adiabatic evolution, the right side of the VHA. Note the significantly better performance of the VHA as compared to the adiabatic evolution. For the data of the rightmost column denoted with $99.90^*$, instead of optimizing the parameters for the faulty gates, we used the parameters as obtained for optimizing with $100\,\%$ fidelity. The comparison demonstrates the capabilities of the VHA in mitigating the errors.
}
\label{tab:adiabatic-vs-vha}
\end{table}

Even for perfect gate fidelities of $100\,\%$ the adiabatic evolution does not improve the state fidelity significantly if we increase $n$. (Note that the initial state fidelity $|\langle \psi_\mathrm{g} | \psi_\mathrm{0} \rangle | = 98.87\,\%$ is already high due to the small system size; no improvement could be achieved for $n=2$.) For a lower gate fidelity of $99.99\,\%$ the adiabatic evolution barely increases the final state fidelity above $99\,\%$. For still lower gate fidelity of $99.9\,\%$ the adiabatic evolution fails to improve the final overlap altogether. This was to be expected from the results of our previous work~\cite{Jan_Noise_Paper}, where we provided estimates for the maximum number of gates that can be handled for a given gate fidelity in a quantum simulation with fixed parameters. Indeed for a gate fidelity of $99.9\,\%$ the gate count for the adiabatic evolution exceeds this limit.

On the other hand, we observe significantly better performance for the VHA. For perfect gates two steps give already a very high final state fidelity; only three steps are necessary to achieve essentially a perfect result (an error of about $10^{-12}$ was observed, which is within numerical inaccuracies). Even with gate errors present we achieve high final state fidelities. The numbers clearly show that introducing more steps helps suppressing the quasi-static errors considered here. 

The rightmost column, labeled by $99.90^*$, illustrates the error mitigation provided by the VHA. For the data in this column we took the optimized parameters for perfect gates and used them in the evolution according to Eq.~\eqref{eq:vha-evolution} with faulty gates with $\overline{\mathcal{F}}_\mathrm{min} = 99.9\,\%$ without any further optimization. This procedure does not take advantage of the potential of the VHA for error mitigation. The low performance of this method illustrates the power of variational error suppression.

The set of gate errors are chosen random but fixed (static) for both methods. However, in different runs they are chosen independent corresponding to the given gate fidelities. The results of Table~\ref{tab:adiabatic-vs-vha} are averaged over many runs with different sets of errors. We can add that the error suppression of the VHA also reduces the standard deviation of the results for different error sets significantly. For the adiabatic evolution and lower gate fidelities we needed of the order of $10^5$ runs in order to reach the shown accuracy of the average. The VHA, on the other hand, needs only a low number of runs for larger $n$, even for low gate fidelities, to achieve the same accuracy, and even a single run is already quite reliable.\footnote{This also alleviates the optimization overhead of the VHA; the algorithm has to be reinitiated hundreds of times in order to optimize the variational parameters.}

\section{Scaling up}

Next we extend the analysis of the variational Hamiltonian ansatz to larger systems. Table~\ref{tab:vha-larger} shows the final state fidelity $\left| \langle \psi_\mathrm{g} | \psi_\mathrm{f} \rangle \right|$ for various step numbers $n$ and gate fidelities $\overline{\mathcal F}_\mathrm{min}$, now for the VHA applied to a $3\!\times\!2$ and a $3\!\times\!3$ Hubbard model. Here we show the results obtained for a single realization of the gate errors for each gate fidelity and step number. Statistical fluctuations are reduced due to the large number of gates per step. Averaging would introduce only small differences to the data presented. It can be ignored, especially for $n > 6$ where the variational error suppression is strong. For the $3\!\times\!2$ and $3\!\times\!3$ systems the algorithm requires 44 and 81 two-qubit gates per iteration step, respectively, i.e., overall up to 810 two-qubit gates were applied to the initial state.

\begin{table}
\begin{tabular}{r r r r r r r}
\hline\hline
\multicolumn{7}{c}{VHA, $3\!\times\!2$}\\
& & \multicolumn{5}{c}{$\bm{\overline{\mathcal{F}}_\mathrm{min}\,[\%]}$}\\
& & \textbf{100.00} & \textbf{99.999} & \textbf{99.990} & \textbf{99.900} & \textbf{99.500}\\
& \textbf{4} & 99.65 & 99.58 & 99.34 & 97.93 & 94.86\\
\multirow{2}{*}{$\bm n$} & \textbf{6} & 99.81 & 99.85 & 99.74 & 98.24 & 97.26\\
& \textbf{8} & 99.98 & 99.91 & 99.61 & 98.98 & 96.12\\
& \textbf{10} & 100.0 & 99.95 & 99.82 & 98.83 & 97.78\\
\hline\hline
\end{tabular}\\[2ex]
\begin{tabular}{r r r r r r r}
\hline\hline
\multicolumn{7}{c}{VHA, $3\!\times\!3$}\\
& & \multicolumn{5}{c}{$\bm{\overline{\mathcal{F}}_\mathrm{min}\,[\%]}$}\\
& & \textbf{100.00} & \textbf{99.999} & \textbf{99.990} & \textbf{99.900} & \textbf{99.500}\\
& \textbf{4} & 99.10 & 98.97 & 98.23 & 95.60 & 86.24\\
\multirow{2}{*}{$\bm n$} & \textbf{6} & 99.59 & 99.46 & 99.27 & 95.55 & 90.06\\
& \textbf{8} & 99.93 & 99.74 & 99.01 & 97.35 & 90.14\\
& \textbf{10} & 99.97 & 99.89 & 99.77 & 98.04 & 90.69\\
\hline\hline
\end{tabular}
\caption{%
Again the final state fidelity $|\langle \psi_\mathrm{g} | \psi_\mathrm{f} \rangle |$ in percent for different values of $n$ and $\overline{\mathcal{F}}_\mathrm{min}$, this time for the variational Hamiltonian ansatz in a $3\!\times\!2$ and a $3\!\times\!3$ system. 
}
\label{tab:vha-larger}
\end{table}

We find again that the VHA produces very high final state fidelities. However, we also notice how high gate fidelities are necessary for a good performance of the algorithm. For a gate fidelity of $\overline{\mathcal{F}}_\mathrm{min} \leq 99.9\,\%$ which should be accessible in the next few years, we did not reach final state fidelities above $99\,\%$.

Further investigations showed that the limited final state fidelities are not necessarily a flaw of the VHA itself but rather of the required optimization. For increasing system size we found the results to be more and more sensitive to the choice of start parameters for the optimization problem. This suggests that the optimizer does not find the global optimum for the parameters but gets trapped in local extrema. The data of Table~\ref{tab:vha-larger} were obtained with a rather limited set of start parameters (see the \hyperref[appx:start-parameters]{Appendix} for further details). 

\begin{table}
\textbf{Improved start parameters}\\
\begin{tabular}{r r r r}
\hline\hline
\multicolumn{4}{c}{VHA, $3\!\times\!2$}\\
& & \multicolumn{2}{c}{$\bm{\overline{\mathcal{F}}_\mathrm{min}\,[\%]}$}\\
		& 			  & \textbf{99.90} & \textbf{99.50}\\
		& \textbf{6}  & 98.91		   & 97.26\\
$\bm n$ & \textbf{8}  & 99.31		   & 98.65\\
		& \textbf{10} & 99.52		   & 98.91\\
\hline\hline
\end{tabular}
\begin{tabular}{r r r r}
\hline\hline
\multicolumn{4}{c}{VHA, $3\!\times\!3$}\\
& & \multicolumn{2}{c}{$\bm{\overline{\mathcal{F}}_\mathrm{min}\,[\%]}$}\\
		& 			  & \textbf{99.90} & \textbf{99.50}\\
		& \textbf{6}  & 96.76		   & 90.81\\
$\bm n$ & \textbf{8}  & 98.02	       & 94.18\\
		& \textbf{10} & 98.91		   & 92.55\\
\hline\hline
\end{tabular}
\caption{%
The final state fidelities in percent after rerunning the VHA with improved sets of start parameters for some of the values for $\overline{\mathcal{F}}_\mathrm{min}$ and $n$ covered in Table~\ref{tab:vha-larger}. One can notice a significant improvement of the final state fidelity.
}
\label{tab:vha-larger-improved}
\end{table}

To substantiate this conclusion we performed the VHA for some values of $\overline{\mathcal{F}}_\mathrm{min} \leq 99.9\,\%$ and $n \geq 6$, exploring a larger set of start parameters (see the \hyperref[appx:start-parameters]{Appendix} for more information). Table~\ref{tab:vha-larger-improved} displays the final state fidelity of the $3\!\times\!2$ and $3\!\times\!3$ systems for the optimized start parameters, showing a significant improvement of the final state fidelity over the results of Table~\ref{tab:vha-larger}. We emphasize that this was not because of a favorable set of random gate errors; once better start parameters were found the results are changing little with varying the gate errors.

Another measure of the performance of VHA is to look at the value of the ground state energy. For the $3\!\times\!3$ Hubbard model with $U=2t>0$ the exact value is $E_\mathrm{g} = -9.67\,t$. For the chosen initial state the state fidelity is already $96.18\,\%$, but the expectation value of the Hamiltonian is only $\langle \psi_0 | H | \psi_0 \rangle = -9.29\,t$. After 10 steps of the VHA for a gate fidelity of $99.90\,\%$ the final state fidelity has improved close to $99\,\%$, and the expectation value of the Hamiltonian reaches $\langle \psi_\mathrm{f} | H | \psi_\mathrm{f} \rangle = -9.60\,t$.

\section{Conclusion}

In this paper we studied in detail the quantum simulation of Hubbard models of small size and with a specific type of gate errors, but our analysis still allows drawing several conclusions:

(i) The variational Hamiltonian ansatz (VHA) produces the ground state wave function of the Hubbard model in good approximation with a number of steps which is much lower than the number of Trotter steps needed in an adiabatic approach.

(ii) The effects of (static) gate errors are strongly mitigated by the variational methods.

(iii) For the considered system size, gate fidelities of the order of $99.9\,\%$,which should be within reach for state-of-the-art digital quantum computers, 
allow preparation of the ground state with a final state fidelity above $99\,\%$.

(iv) This performance can be reached with a low number of variational parameters per step (5 in our case for the 2D Hubbard model).

It is clear that introducing more variational parameters, up to one parameter per gate, would enhance the variational error suppression of quasi-static errors. Eventually it leads to approaches like the variational quantum eigensolver~\cite{VQE_Paper}. However, introducing more parameters poses a challenge to the classical optimization routines. We found that even for our small set of parameters, the emerging optimization problem poses a substantial obstacle, since with growing system size the gradients with respect to the variation of the parameters decrease~\cite{Inital_Guess_Problem}. The difficulties with the optimization algorithms appear a stronger limitation of the performance of variational algorithms than a limited set of variational parameters.

Better optimization algorithms could help with further issues. One could consider non-static portions of gate errors and statistical measurement errors. Such fluctuating errors are difficult to manage for optimizers, particularly for gradient based optimization protocols. We also noted the need for a good choice of the initial set of variational parameters, as well as of the initial state. The latter could be obtained, e.g., from mean field theory. Finally more advanced quantum gate sequences implementing the terms of the Hamiltonian can lower the gate count and reduce the impact of gate errors. We applied up to 810 two-qubit gates in our examples of rather small systems. Algorithms with superior scaling behavior for Hubbard models~\cite{Good_Algorithm,Efficent_Hubbard} should be considered to improve the performance of the VHA for larger systems.

\section{Appendix}

\appendix

\section{Gate sequence}
\label{appx:gate-sequence}

For illustration we show the gate sequence producing the unitary transformation $\e^{\mathrm{i} \theta_{\alpha,k} H_\alpha}$ from Eq.~\eqref{eq:vha-evolution}. The different $H_\alpha$, introduced in the main text, contain either hopping terms $-t (c^\dagger_{j,s} c^{\phantom \dagger}_{j',s} + c^\dagger_{j',s} c^{\phantom \dagger}_{j,s})$ or on-site interactions $U c^{\dagger}_{j,\uparrow} c^{\phantom \dagger}_{j,\uparrow} c^\dagger_{j,\downarrow} c^{\phantom \dagger}_{j,\downarrow}$. We assume that the hardware of the quantum computer allows for $ZZ$-like and $XX+YY$ interaction and, for simplicity, unrestricted connectivity between the qubits. The terms summarized in each of the $H_\alpha$ commute among each other making their ordering irrelevant.

For the following discussion it is convenient to absorb the spin index in a consecutive numbering of the lattice sites via $(j,\uparrow) \mapsto j$ and $(j,\downarrow) \mapsto j+M$ where $M$ is the total number of sites. With this notation the Jordan-Wigner transformation becomes $c_j = \prod_{l=1}^{j-1} (-\sigma^z_l) \sigma^-_j$, which involves the \emph{Jordan-Wigner string} $\prod_{l=1}^{j-1} (-\sigma^z_l)$.

The on-site terms now read $U \sp j \sm j \sp{j+M} \sm{j+M}$. In the variational approach the gate $\e^{\mathrm{i} \theta U \sp j \sm j \sp{j+M} \sm{j+M}}$ needs to be implemented with some parameter $\theta$, which embodies a $ZZ$ interaction (up to some single-qubit phases). To account for the gate errors we add an over-rotation $\delta \theta$ to the parameter, the size of which depends on the assumed gate fidelity.

\begin{figure}
\centering
\includegraphics[width=\columnwidth]{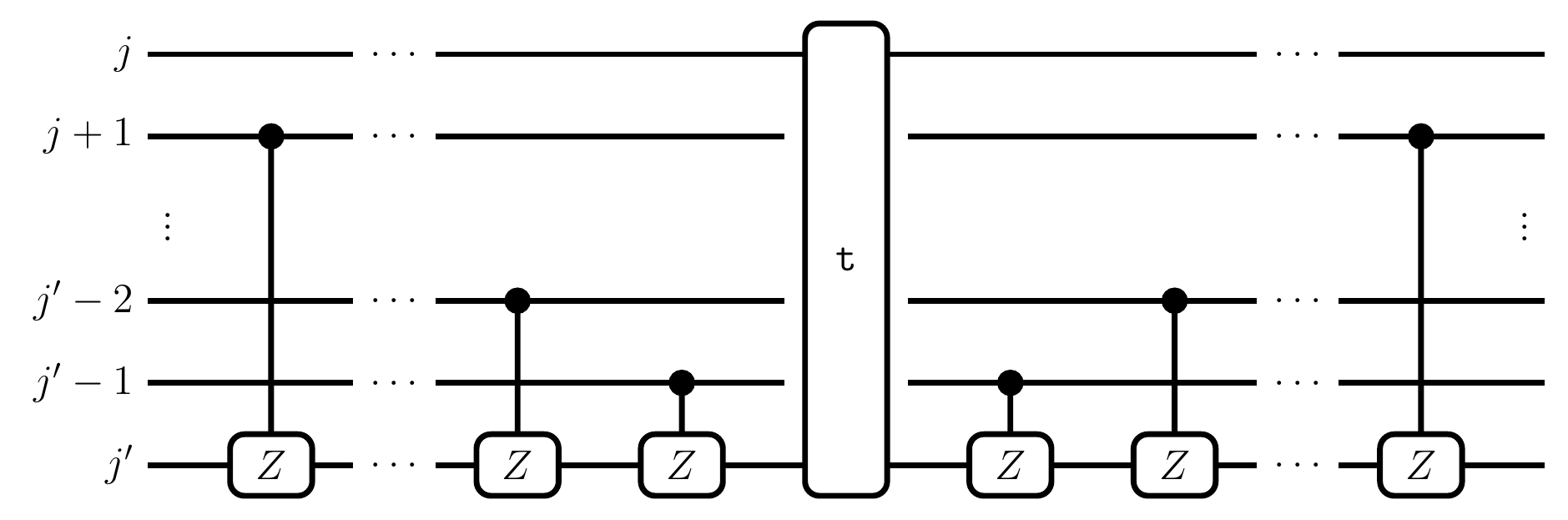}
\caption{%
A \gate t gate implementing an $XX+YY$ interaction between qubits $j$ and $j'$ is nested between controlled $Z$ gates. The \gate{CZ} gates introduce a Jordan-Wigner string such that the gate sequence implements a fermionic hopping term between the orbitals $j$ and $j'$.
}
\label{fig:hopping-sequence}
\end{figure}

The hopping terms (where, say, $j<j'$) transform to $-t \big(\sp j \sm {j'} \, \prod_{l=j+1}^{j'-1} (-\sigma^z_l) + \mathrm{H.c.}\big)$. We do not assume the hardware to allow for more than two-qubit interactions. A hopping term, can be modeled by the $XX+YY$ interaction, i.e., $\e^{-\mathrm{i} \theta t (\sp j \sm{j'} + \sp{j'} \sm j)}$, which we denote as \gate t gate. Again, gate errors lead to an over-rotation $\delta \theta$ to be added to the parameter $\theta$. Residual Jordan-Wigner strings can then be implemented by sandwiching the \gate t gate with controlled $Z$ gates (\gate{CZ}) as displayed to Fig.~\ref{fig:hopping-sequence}. The \gate{CZ} gate has again $ZZ$-like interaction, and we implement it as $\e^{\mathrm{i} \pi \sp j \sm j \sm {j'} \sp {j'}}$ for the control qubit $j$ and the target qubit ${j'}$. Over-rotations are introduced as an addition $\delta \varphi$ to the angle $\pi$.

\section{Choice of start parameters}
\label{appx:start-parameters}

When using the VHA, in order to minimize the energy according to Eq.~\eqref{eq:vha-energy}, one has to start with an initial guess for the parameters $\vec \theta$. In Table~\ref{tab:adiabatic-vs-vha} and Table~\ref{tab:vha-larger} for each value of $n$ and $\overline{\mathcal{F}}_\mathrm{min}$ we tried three different sets of start parameters, motivated by some physical reasoning:

1. Since the VHA is inspired by the adiabatic evolution one of our choices of parameters $\theta_{\alpha,k}$ from Eq.~\eqref{eq:vha-evolution} was to mimic the adiabatic evolution. We set $\theta_{\alpha,k} = \frac{1}{t}$ for $\alpha \in \{1, \dots, 4\}$ (i.e., the hopping elements) and $\theta_{5,k} = \frac{k}{n} \frac{1}{t}$ (i.e., the interaction terms), with $t$ being the hopping energy of the Hamiltonian~\eqref{eq:Hamiltonian}. This represents the adiabatic evolution~\eqref{eq:adiabatic-evolution} during time $\tau = \frac{n}{t}$.

2. We chose a parameter set where not only the interaction but also the hopping is turned on gradually, i.e., $\theta_{\alpha,k} = \frac{k}{n} \frac{1}{t}$ for all $\alpha$.

3. We noticed that the optimized parameters of the VHA usually do not resemble an adiabatic path or show steady growth, rather they are more evenly distributed. Hence, we chose a set where $\theta_{\alpha,k} = \frac{1}{n} \frac{1}{t}$ for all $\alpha$. This set represents a Trotter expansion with $n$ steps for a time evolution for the duration $\tau = \frac{1}{t}$.

For all choices of initial parameters, after the optimization the final values were vastly different. We conclude that for the larger systems one often gets stuck in a local optimum. For this reason, we tried further sets of initial parameters with results shown in Table~\ref{tab:vha-larger-improved}:

Firstly, we added initial parameters recreating an adiabatic evolution similar to point $1$ described above, but for $\tau = \frac{1}{t}$. Secondly, we chose an even distribution, similar to point $3$, but with $\theta_{\alpha,k} = \frac{r}{t}$ where we varied $r \in \{0.1, 0.2, \ldots, 1.0\}$. (We noticed that usually the magnitude of the final parameters were between zero and $\sim \frac{1}{t}$.) These additional start parameters improved the data of Table~\ref{tab:vha-larger} towards the results of Table~\ref{tab:vha-larger-improved}.

A deeper understanding how to deterministically find suitable initial parameters should be acquired. However, this is out of the scope of this work. For this paper the tested parameter sets already helped to show the capabilities of the VHA, achieving remarkable results in the considered small systems.

%

\end{document}